\documentclass[twocolumn,showpacs,preprintnumbers,prb,fleqn]{revtex4}
\usepackage{epsfig}
\usepackage{graphicx}
\usepackage{amsfonts}
\newcommand{\wt}{\widetilde}
\begin{document}
                                                                                

\title{Critical behaviour of mixed random fibers, fibers on a chain and random 
graph}
\author{Uma Divakaran}
\email{udiva@iitk.ac.in}
\author{Amit Dutta}
\email{dutta@iitk.ac.in}
\affiliation{Department of Physics, Indian Institute of Technology Kanpur - 208016}
\date{\today}
\begin{abstract}
We study random fiber bundle model (RFBM) with different threshold 
strength distributions and load sharing rules. A mixed RFBM within 
global load sharing scheme is introduced
which consists
of weak and strong fibers with uniform distribution of threshold strength of 
fibers having a discontinuity. 
The dependence of the critical stress of the above model on the 
measure of the discontinuity of the
distribution  is extensively studied. 
A similar RFBM with two types of fibers belonging to two different Weibull 
distribution of 
threshold strength
is also studied.  The variation of the critical stress of a one dimensional
RFBM with the number of fibers is obtained for strictly uniform distribution
and local load sharing  
 using an exact
 method which assumes one-sided load transfer. The critical behaviour of 
RFBM with fibers placed on a
random graph having co-ordination number 3 
is investigated numerically for uniformly distributed threshold strength of fibers subjected to local load sharing rule, and
 mean field critical behaviour is established.

\end{abstract}
\maketitle

\section{Introduction}
Sudden catastrophic failure of structures due to unexpected fracture of 
component materials is a concern and a challenging problem of physics as well
as engineering.
The dynamics of the failure of materials show interesting properties and hence 
there has
been an enormous amount of study on the breakdown phenomena till now\cite{Benguigui,Sahimi,Herrmann,Bak,Rava}. 
A total analysis of the fracture of materials should include
the investigation of a possible fracture before the occurrence of large 
deformations and hence the determination of the
critical load beyond which if the load on the system increases, 
the complete failure occurs. Future progress in the understanding of 
fracture and the application of that knowledge demands the integration 
of continuum mechanics with the scientific disciplines of material science,
 physics, mathematics and chemistry. The main challenge here is to
 deal with the inherent stochasticity of the system and the statistical 
evolution of microscopic crack to accurately predict the point of final 
rupture.

Having said about the complexity involved in fracture processes, one can still 
get an idea of fracture phenomena by studying simplified models. The simplest
available model is Fiber Bundle Model(FBM)\cite{Daniel, Coleman, Peirce}. 
It consists of N parallel
fibers. All of them have their ends connected to horizontal rods as shown in 
Fig.~1. The disorder of a real system is introduced in the fiber bundle in the 
form of random distribution of strength of each fiber taken from a probability 
distribution P($\sigma$) and hence called Random Fiber Bundle Model (RFBM). 
The strength of each fiber is called its threshold 
strength. 
As a force F is applied externally on a bundle of N fibers, a stress 
$\sigma=F/N$
acts on each of them. The fibers which have their threshold strength
smaller than the stress generated, will break immediately. The next question 
that arises is the affect of breaking of fibers on the remaining intact 
fibers $i.e.$, one has to decide a load sharing rule. 
The two extreme cases of load
sharing mechanisms are Global Load Sharing (GLS)\cite{Daniel,dynamic} 
and Local Load Sharing (LLS)\cite{Pacheco, Zhang,wu}.
In GLS, the stress of the
broken fiber is equally distributed to the remaining intact fibers. This rule
neglects local fluctuations in stress and therefore is effectively a 
mean field model
with long range interactions among the elements of the system \cite{Stanley}.
On the other hand, in LLS the stress of the broken fiber is given only 
to its nearest surviving neighbours. It is obvious that the actual breaking process
involves a sharing rule which is in between GLS and LLS. Several studies
have been made which considers a rule interpolating between GLS and LLS
\cite{mixed1,mixed2}.

\bigskip
\begin{figure}
\includegraphics[height=1.5in]{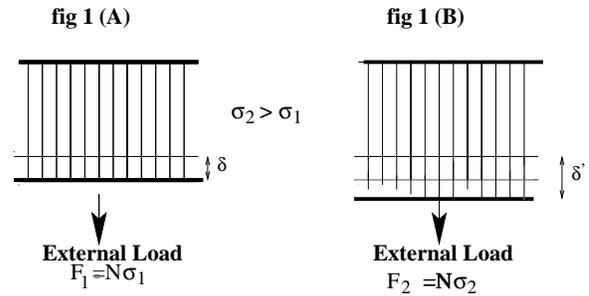}
\caption {The vertical parallel lines correspond to fibers which are connected 
to two horizontal plates (shown with bold lines). The upper plate is fixed
and the lower plate is movable. Fig.~1(a) shows a smaller force $F_{1}$ applied
on the bundle producing a strain $\delta$. Fig.~1(b) corresponds to a higher 
value of load, leading to a strain $\delta^{'}$ and fibers begin to break.}
\end{figure}
For a given force F, the fibers break and distribute their load 
following a load sharing rule until all
the fibers have their threshold strength greater than the redistributed stress
acting on them. This corresponds to the fixed point of the 
dynamics of the system.
As the applied force is increased on the system, more and more fibers break
and the additional stress is redistributed amongst the other fibers. There 
exist a critical load (or stress $\sigma_{c}$) beyond which if the load is 
applied, complete failure of the system takes place. Most of the studies on
FBM involve the determination of the critical stress $\sigma_{c}$ and the 
investigation of the type 
of phase transition from a state of partial failure to a state of complete 
failure. It has been shown that a bundle following GLS has a finite value of critical stress and belongs to a universality class with a 
specific set of critical exponents\cite{dynamic,pratip} 
whereas there is no finite critical stress $\sigma_{c}$ at thermodynamic 
limit in the case of LLS in one dimension\cite{Pacheco,Hansen,Smith,reviewchak}.
On the other hand, LLS on complex network has been shown to belong to the 
same universality
 class as that of GLS with the same critical exponents\cite{complex}.

The paper is organised as follows. Section II consists of mixed fibers with 
uniform distribution (UD) of threshold strength of fibers and GLS. 
The critical behaviour of such a model is 
explored analytically. Mixed fibers with Weibull distribution (WD) 
of threshold 
strength along with GLS is studied using a quasistatic approach of load 
increase in section III. One dimensional, one sided LLS with strictly 
uniform distribution (UD)
is presented in section IV. Section V consists of a bundle of fibers
whose tips are placed on a random graph with co-ordination number 3.
The main conclusions are presented in section VI.

\section{Mixed Fibers with Uniform Distribution and GLS}
Let us consider a $'\rm mixed'$ fiber bundle model with normalised density
distribution of threshold strength as shown below:
\begin{figure}[h]
\hspace{1.0in}\includegraphics[height=2.0in,width=2.5in]{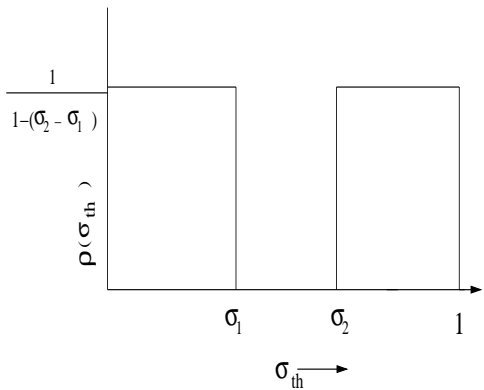}
\caption  {Mixed Uniform Distribution.}
\end{figure}

\begin{eqnarray}
\rho(\sigma_{th}) &=& \frac {1}{1-(\sigma_{2}-\sigma_{1})}~~~~~~
0<\sigma_{th}\leq \sigma_{1}\nonumber\nonumber\\
&=&0~~~~~~~~~~~~~~~~~~~~~~~\sigma_{1}<\sigma_{th}<\sigma_{2}\nonumber\\
~~~~~~~~~~&=&\frac {1}{1-(\sigma_{2}-\sigma_{1})}~~~~~~~\sigma_{2}\leq\sigma_{th}\leq 1. 
\end{eqnarray}
The present model clearly consists of fibers of two different types.
This classification is based
on whether a randomly chosen fiber from the bundle belongs to class A of fibers
($0<\sigma_{th}\leq \sigma_{1}$) or class B of fibers ($\sigma_{2}\leq\sigma_{th}\leq 1$)
as shown in Fig.~2. In other words, the model consists of weak and strong
fibers but no fibers with strengths between $\sigma_{1}$ and
$\sigma_{2}$, which is the forbidden region for the fiber strength.
Since we are considering GLS here, the exact location of class A or class B
fibers is immaterial.
The aim is to determine the critical stress $\sigma_{c}$ for this model
such that if the external stress exceeds $\sigma_c$ , the bundle breaks 
down completely.

If a fraction $x$ of the total fibers belong to class A and the 
remaining $1-x$ to class B , then to maintain the uniformity of the distribution(Eq.~(1)), we must have
\begin{eqnarray}
\frac{1}{1-(\sigma_{2}-\sigma_{1})}\int_{0}^{\sigma_{1}} d\sigma = x\nonumber\\
\frac{\sigma_{1}}{1-(\sigma_{2}-\sigma_{1})} = x\nonumber\\
\noindent {\rm or},~~\sigma_{1} = \frac{x}{1-x}(1-\sigma_{2}).
\end{eqnarray}
The above equation provides the relationship between $x$, $ \sigma_1$ and 
$\sigma_2$ showing that once $x$ and $\sigma_1$ are set, $\sigma_2$ is
fixed automatically. This equation at the same time puts a restriction
on $\sigma_1$. One can check that if $\sigma_1>x$, $\sigma_2$
becomes smaller than $\sigma_1$, which is impossible with the given form
of distribution (Eq.~(1)). We shall therefore restrict the value of
$\sigma_1$ to be smaller than $x$.

The mechanism of the failure of RFBM is the following:
under an external stress $\sigma$, a fraction of fibers
having threshold strength less than the applied stress  
fail immediately. The additional load due to breaking
of these fibers is redistributed globally among the remaining intact fibers.
 This redistribution causes further failures.
Let us define $U_{t}=N_{t}/N$ as the fraction of unbroken fibers after a time
step t. Then,

$$\sigma_{t}=\frac{F}{N_{t}}=\frac{\sigma}{U_{t}}.$$

The recurrence relation between $U_{t}$ and $U_{t+1}$ for a given applied
stress $\sigma$
is obtained as \cite{dynamic}:
\begin{eqnarray}
U_{t+1}=1-P(\sigma_{t})=1-P(\frac{\sigma}{U_{t}}).
\end{eqnarray}
This dynamics propagate (in discrete time) until no further breaking takes
place.

If the initial applied load is very small so that the redistributed 
stress is always less than $\sigma_{2}$, then no fibers from class B will fail. Thus for breaking of class B fibers, the redistributed stress has to be greater than $\sigma_{2}$. 
Let $\sigma_{t}$ (the redistributed stress) cross
$\sigma_{2}$ at some instant t. Then the critical stress of 
the system can be calculated as follows:
\begin{eqnarray}
U_{t+1} &=& 1-P(\frac{\sigma}{U_{t}}) \nonumber \\
 &=& 1-[\frac{\sigma_{1}}{1-(\sigma_{2}-\sigma_{1})} +\frac{1}{1-(\sigma_{2}-\sigma_{1})} (\frac{\sigma}{U_{t}}-\sigma_{2})]\nonumber.
\end{eqnarray}

At fixed point of the dynamics, when no further breaking takes place, 
$U_{t+1}=U_{t}=U^{*}$ and
we get,

$$ U^{*}= 1-[\frac{\sigma_{1}}{1-(\sigma_{2}-\sigma_{1})} +\frac{1}{1-(\sigma_{2}-\sigma_{1})} (\frac{\sigma}{U^{*}}-\sigma_{2})] $$
giving,
\begin{eqnarray}
U^{*} &=& \frac {1}{2(1-(\sigma_{2}-\sigma_{1}))} [1 \pm \sqrt
{1-4[1-(\sigma_{2}-\sigma_{1})]\sigma}] \nonumber \\
U^{*} &=& \frac {1}{2(1-(\sigma_{2}-\sigma_{1}))} [1 \pm \sqrt
{1-\frac{\sigma}{\sigma_{c}}]}
\end{eqnarray}
 where
\begin{eqnarray}
\sigma_{c}=\frac {1} {4[1-(\sigma_{2}-\sigma_{1})]}
\end{eqnarray}
It can be shown easily that the solution with $+$ in the parenthesis
gives a stable
solution. Therefore,  $\sigma_{c}$ is the critical stress of the system because
 if the applied
stress $\sigma$ is less than $\sigma_{c}$, the system reaches a fixed point.
If the
applied stress is greater than $\sigma_{c}$, $U^{*}$ becomes imaginary
and the bundle breaks down completely. It is interesting to note that
at $\sigma=\sigma_{c},~U^{*}=1/2[1-(\sigma_{2}-\sigma_{1})]$  $i.e.$,  
more than half of the fibers are still unbroken at critical point. 

We should now look at the restrictions on $\sigma_1$, $\sigma_2$ and $x$
(other than $\sigma_1 <x$) so that the above calculations are valid.  
From Eq.~(4), we find that if the width of the forbidden region ($\sigma_2 -\sigma_1$) is greater than half, $U^{*}(\sigma_c )>1$ which is unphysical.
Therefore, $(\sigma_2 - \sigma_1)$ should be smaller than 1/2 for all values
of $x$ for the above calculations to hold. Using this restriction in 
Eq.~(5), we find that the critical stress $\sigma_c$ will always be less than 
1/2. Now, looking at the physical meaning of $U^*$, we must also check that the
fraction of broken fibers at the fixed point corresponding to critical stress
$\sigma_{c}$ must be greater than the
fraction of fibers having their threshold stress smaller than the critical
stress. We summarise all the restrictions below:

\begin{equation}
\sigma_1 \leq x, ~~  \sigma_2 - \sigma_1 < 0.5 ~~\rm and~~(1- U^*(\sigma_c)) > P(\sigma_c)  
\end{equation}

In the table below, we enlist some allowed values of $\sigma_1$, $\sigma_2$ for
different $x$ and also the corresponding critical stress.
\begin{center}
\begin{tabular}{|c|c|c|c|}\hline
~~~~x~~~~&~~~~ $\sigma_{1}$~~~~  &~~~~ $\sigma_{2}$ ~~~~ & ~~~~$\sigma_{c}$\\\hline
0.10  &0.08  &0.28  &0.31\\\hline
0.20  &0.19  &0.24  &0.26\\\hline
0.30  &0.25  &0.42  &0.30\\\hline
0.30&0.29&0.33&0.26\\\hline
0.40&0.35&0.47&0.29\\\hline
0.50&0.45&0.55&0.27\\\hline
0.80&0.7&0.82&0.28\\\hline
\end{tabular}
\end{center}
From the table, we observe that
the critical stress depends entirely on the width of the forbidden region,
$i.e.$, $\sigma_2 - \sigma_1$.  Also, 
when $\sigma_{1}$ is increased for a fixed $x$, $\sigma_{2}$
decreases and hence $\sigma_{c}$ decreases. Physically, decreasing 
$\sigma_2$  (with $x$ fixed) means that
lowest threshold of the  strong fibers (class B)  gets lowered. 
It is found that $\sigma_{1}$
cannot be very small compared to x, else all the above mentioned 
conditions do not get satisfied.

One can define an order parameter as shown below:

\begin{eqnarray}
2[1-(\sigma_{2}-\sigma_{1})]U^{*} - 1 = (\sigma_{c}-\sigma)^{\frac{1}{2}}=
(\sigma_{c}-\sigma)^{\beta}
\end{eqnarray}
The order parameter goes to 0 as $\sigma \rightarrow \sigma_{c}$
following a power law  $(\sigma_{c}-\sigma)^{\frac{1}{2}}$ . 
Susceptibility can be defined as the increment in the number of broken fibers
for an infinitesimal increase of load. Therefore,
\begin{eqnarray}
\chi=\frac{dm}{d\sigma}~~~ {\rm where}~~~ m = N[1-U^{*}(\sigma)]\nonumber
\end{eqnarray}
Hence,
\begin{eqnarray}
\chi \propto (\sigma_{c}-\sigma)^{-\frac{1}{2}}=(\sigma_{c}-\sigma)^{-\gamma}
\end{eqnarray}

\begin{figure}[h]
\includegraphics[height=2.0in,width=2.1in]{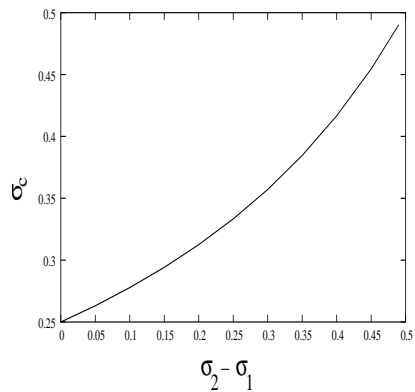}
\caption  { Variation of $\sigma_{c}$ with $\sigma_{2}-\sigma_{1}$ for a
mixed RFBM with a distribution as in Fig.~2}
\end{figure}
As expected, the model follows mean field exponents with $\beta=1/2$
and $\gamma=1/2$.
The interesting features of this model are: In this model, the forbidden
zone has a nontrivial role to play. (i)By changing $(\sigma_{2}-\sigma_{1})$ 
and x,
we can tune the strength of the system. However there are restrictions on
the parameters as discussed above.
$\sigma_{c}$ as a function of 
$(\sigma_{2}-\sigma_{1})$ is shown in Fig.~3.
(ii) If x = 0, $\sigma_{2}\ne 0, \sigma_{1}=0 (\rm ~from~ Eq.~ 2)$ which gives
$\sigma_{c}=1/4(1-\sigma_{2})$ and we observe an elastic to 
plastic deformation as discussed in reference 9. 
In this case, for $\sigma<\sigma_{2}$, 
the fiber in the bundle obey Hook's law. As soon as $\sigma>\sigma_{2}$, 
the dynamics of failure of fibers start.
(iii)If $\sigma_{2}-\sigma_{1}$=0, the problem is that of a simple UD
with threshold strength between 0 and 1 which has a critical stress
$\sigma_{c} = 0.25$ \cite{dynamic}. 

These results confirm that the failure of mixed fiber bundle model
with Global Load Sharing belongs to a universality class with critical
exponents $\beta=1/2$ and $\gamma=1/2$.

\section{Mixed Fibers with Weibull Distribution and GLS}

In the Weibull distribution of threshold strength of fibers, the 
probability of failure of each element has the form
\begin{equation}
P(\sigma)=1-e^{-(\frac{\sigma}{\sigma_{0}})^{\rho}}
\end{equation}
where $\sigma_{0}$ is a reference strength and $\rho$ is the Weibull 
index. We now consider a mixed RFBM as in section II with WD
of threshold strength of fibers. 
As in the UD case (of Sec.~(2)), here also a fraction $x$ of fibers belong
to the class A (WD with $\rho=2$) and the remaining $(1-x)$ fraction belong
to the class B (WD with $\rho =3$)  
with the reference strength $\sigma_{0}$ equal to 1 in either case.
 
For the conventional WD, 
consider  a situation where the stress on each fiber increases from 
$\sigma_{1}$ to $\sigma_{2}$.  The probability that a fiber randomly chosen 
from the WD survives from the load $\sigma_{1}$ but fails 
when the load is $\sigma_{2}$, is given by
$$p(\sigma_{1},\sigma_{2})=\frac{P(\sigma_{2})-P(\sigma_{1})}{1-P(\sigma_{1})}=
1-e^{-(\sigma_{2}^{\rho}-\sigma_{1}^{\rho})}.$$
Thus the probability that the chosen fiber that has survived the load $\sigma_{1}$ also survives the (higher) load $\sigma_{2}$ is 
$q(\sigma_{1},\sigma_{2})=e^{-(\sigma_{2}^{\rho}-\sigma_{1}^{\rho})}$.
The key point is that the force F on the bundle is increased quasistatically
so that only one fiber amongst the remaining intact fibers break.
As mentioned previously, the dynamics of breaking will continue 
till the system reaches a fixed point. The process of slow increase of external
load is carried on up to the critical stress $\sigma_{c}$.

 To deal with the nature of the present
$'\rm mixed'$ random fiber bundle model, the method used by Moreno, Gomez 
and Pacheco\cite{mixed3} is generalised in the following manner.
It is important here to keep track of number of unbroken fibers in each 
distribution separately.
Let us assume that after a loading is done and a fixed point is reached, 
$N_{k_{2}}$ and $N_{k_{3}}$ are the number of unbroken fibers corresponding to 
$\rho=2$ and $\rho=3$ distribution respectively, where each fiber has a stress $\sigma_{k}$.  
\begin{figure}[h]
\includegraphics[height=2.2in,width=2.1in]{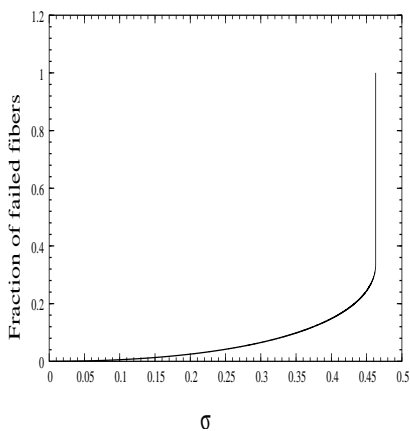}
\caption{Variation of fraction of broken fibers with external load
per fiber  $\sigma$ for a RFBM with two distributions corresponding 
to $\rho=2 ~\rm and ~\rho=3$. N=50000, $x$=0.5}
\end{figure}

We define $q_{2}$ and $q_{3}$ as follows:
\begin{eqnarray}
q_{2}(\sigma_{1},\sigma_{2}) = e^{-(\sigma_{2}^{2}-\sigma_{1}^{2})}\nonumber\\
q_{3}(\sigma_{1},\sigma_{2}) = e^{-(\sigma_{2}^{3}-\sigma_{1}^{3})}\nonumber.
\end{eqnarray}
The load $(N_{k_{2}}+N_{k_{3}})\sigma_{l}$ that has to be applied to break
one fiber is given by $\sigma_{l}=min(\sigma_{l_{2}},\sigma_{l_{3}})$
where $\sigma_{l_{2}}$ (or $\sigma_{l_{3}}$) is the next weakest fiber
in the $\rho=2$ (or $\rho=3$) distribution of strength of fibers and is
obtained by the solution of the following equations.
\begin{eqnarray}
N_{k_{2}}-1 = N_{k{2}}q_{2}(\sigma_{k},\sigma_{l_{2}})\nonumber\\
\noindent {\rm and}~~N_{k_{3}}-1 = N_{k_{3}}q_{3}(\sigma_{k},\sigma_{l_{3}})\nonumber
\end{eqnarray}
which gives
\begin{eqnarray}
\sigma_{l_{2}} = [\sigma_{k}^{2}-{\rm ln}(1-\frac{1}{N_{k_{2}}})]^{1/2}\\
\sigma_{l_{3}} = [\sigma_{k}^{3}-{\rm ln}(1-\frac{1}{N_{k_{3}}})]^{1/3}
\end{eqnarray}
The breaking of one fiber and the redistribution of its stress causes some
more failures. Let us say that during this avalanche, at some point before 
the fixed point is reached, there are $\wt{N_{k_{2}}}$ and $\wt{N_{k_{3}}}$ 
number of unbroken fibers belonging to the two distributions where 
each fiber is under a stress $\wt{\sigma_{k}}$. This stress causes 
some more failures and as a result $N_{k_{2}}^{'}$ and $N_{k_{3}}^{'}$ fibers 
are unbroken. Let the new stress developed is $\sigma_{k}^{'}$.
The number of fibers which survive $\wt{\sigma_{k}}$ and $\sigma_{k}^{'}$
are obtained using the relation
\begin{eqnarray}
N_{k_{2}}^{''} = N_{k_{2}}^{'}q_{2}(\wt{\sigma_{k}},\sigma_{k}^{'})\\
N_{k_{3}}^{''} = N_{k_{3}}^{'}q_{3}(\wt{\sigma_{k}},\sigma_{k}^{'})
\end{eqnarray}
The stress on each fiber is now equal to 
$(N_{k_{3}}^{'}+N_{k_{2}}^{'})\sigma_{k}^{'}/(N_{k_{2}}^{''}+N_{k_{3}}^{''})$.
Eq (12) and Eq. (13) are used again and again until a fixed point is reached.
The fixed point condition is given by 
$(N_{k_{2}}^{'}+N_{k_{3}}^{'})-(N_{k_{2}}^{''}+N_{k_{3}}^{''})<\epsilon$ 
where $\epsilon$ is a small 
number (0.001). However, the critical behaviour does not depend on 
the choice of $\epsilon$.

After the fixed point is reached, stress $\sigma_{l}$ is calculated once again as mentioned before and the whole process is repeated till complete failure
 occurs at $\sigma_{c}$. The above mentioned approach has the advantage 
that it does not require the random averaging of Monte Carlo simulations. 

Fig.~4 shows the fraction of total number of broken fibers as 
a function of applied stress $\sigma$ for the case $x=0.5$. 
The graph clearly shows the existence
of a critical stress, $\sigma_c = 0.46$, which lies between the critical stress 
of $\rho=2$ (0.42) and $\rho=3$ (0.49) distribution .
These theoretical values are obtained from the analytical
expression \cite{Daniel}$\sigma_{c}=(\rho e)^{-1/\rho}$.
The avalanche size S is defined as the number of broken fibers between
two successive loadings.
It diverges near the critical point with an exponent 
$\gamma=1/2$ as $(\sigma_{c}-\sigma)^{-\gamma}$. Scaling behaviour of avalanche 
size S is shown in Fig.~5. This confirms the mean field nature of GLS models
even with mixed fibers. The most attractive feature of the mixed RFBM is
shown in Fig.~(6) which contains the variation of $\sigma_c$ with the fraction
$x$ of class A fibers. Interestingly, $\sigma_c$ decreases linearly as $x$ 
is increased.
\begin{figure}[h]
\includegraphics[height=2.2in,width=2.1in]{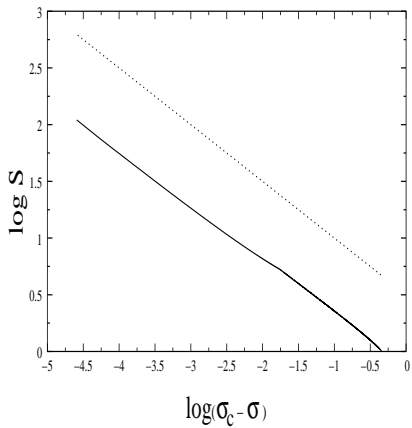}
\caption{Scaling behaviour of avalanche size as the critical point
is reached for the mixed model. Also shown is a straight line (dotted) 
with slope (-1/2). Here, N=50000 and $x=0.5$.}
\end{figure}

\begin{figure}[h]
\includegraphics[height=2.2in,width=2.1in]{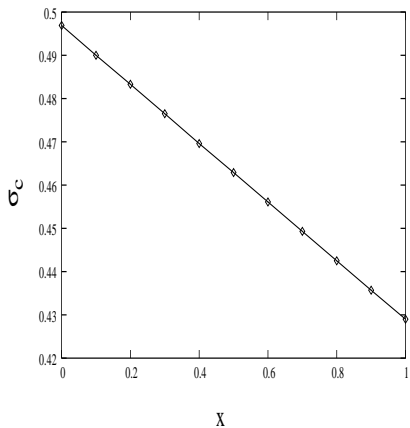}
\caption{Variation of critical stress $\sigma_{c}$ with $x$. Slope of
the straight line is 0.068. N=50000 }
\end{figure}
\section{One Dimensional, One sided LLS}
In numerical studies, it is difficult to generate a strictly uniform
distribution of the threshold strength of random fibers. In this
work, we have addressed the problem of random fibers with UD of threshold 
with LLS scheme  using an exact treatment of one sided load 
distribution\cite{Pacheco}. 
The complication arising in the present approach due to the uniform 
nature of the threshold distribution will also be emphasised later.

In one sided LLS, the additional load due to breaking of a fiber is transferred
only to its neighbour in a 
given direction. The fibers are placed along the circumference 
of a circle with periodic boundary conditions. The method is based 
on the $k$-failure concept. If $k$ is chosen appropriately, 
then more than $k$ adjacent failures 
is equivalent to complete failure of the system. The scheme 
involved is as follows: let $p_{i}=p(\it{i\sigma}$) is the probability 
of failure
of a fiber supporting a load $i\sigma$ (where $\sigma$ is the initial 
applied load per fiber). This is same as the probability of failure 
of all the previous $(\it{i}-1)$ fibers.
Let $P_{N}(\sigma)$ represents the probability
that a system of $N$ fibers break when an initial force equal to $N\sigma$ is 
applied and $Q_{N}(\sigma)=1-P_{N}(\sigma)$ is the probability
that the bundle does not break because of initial load $N\sigma$. We have 
defined the critical load $\sigma_{c}$ by setting $P_{N}(\sigma_{c})=0.5$. 
However, any other choice of $P_{N}(\sigma_{c})$ does not alter the result.
$Q_{N}(\sigma)$ can be easily identified as the probability that out of N
fibers, there is no chain of adjacent broken fibers of length greater than $k$.

$Q_{N}(\sigma)$ can be 
calculated by using a simple iterative procedure described in Fig.~7.
\begin{figure}[h]
\includegraphics[height=2.4in]{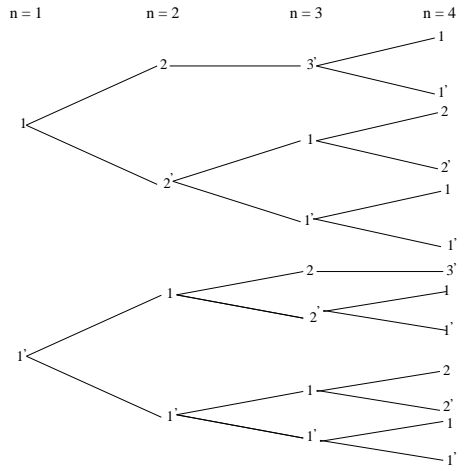}
\caption{Possible breaking configurations of a set with N=4 elements when
the maximum allowed crack length is k=2 for one sided one dimensional LLS}
\end{figure}
The figure shows all possible states of a fiber in a system of 4 fibers
 with maximum number of allowed
adjacent
 broken fibers equal to 2 (i.e., $k=2$). The notation means the following: the
symbols 1 and 2 mean that in the corresponding position, the fiber breaks with 
a stress $1\sigma$, $2\sigma$ respectively. $1',2',3'$ means that the 
corresponding element does not break under the stress $1\sigma$, $2\sigma$ and
$3\sigma$ respectively. This simple notation can be extended to a system 
of $N$ fibers. If a fiber site has a symbol $m$, the two possibilities for
the next fiber are $(m+1)$ and $(m+1)'$. If the site has $m'$, 
the two possibilities are $1$ and $1'$ for the next site as $m'$ 
means that the fiber has not broken 
and hence no 
additional stress is given to the next fiber. But if $m=k$, then the 
only allowed possibility for the fiber at the next site is $(k+1)'$. 
This is to ensure that
we are looking at only those configurations in which the complete breakdown
does not take place. The aim is to write $Q_{N}(\sigma)$ in terms of 
$Q_{N-1}(\sigma)$. Let $B_{i}(N)$, where $i\leq k$ is the sum of 
probability contributions of the paths of length $N$ ending in the symbol i.
Also, $B_{i'}(N)$ with $i'\leq k+1$, is the sum of contributions of all paths 
ending in symbol $i'$. The following relations can then be obtained 
\begin{eqnarray}
&&B_{1}(N)=p_{1}\sum_{i=1}^{k+1} {B_{i^{'}}}(N-1)\\ \nonumber
&&B_{i}(N)=p_{i}B_{i-1}(N-1), ~~1< i \leq k  \\ \nonumber
&&B_{1^{'}}(N)=q_{1}\sum_{i=1}^{k+1} B_{i^{'}}(N-1)\\ \nonumber
&&B_{i^{'}}(N)=q_{i}B_{i-1}(N-1),~~~1< i^{'} \leq k+1.
\end{eqnarray}
Thus $Q_{N}$ can be defined as
\begin{equation} ~~~~Q_{N}=\sum_{i=1}^{k} B_{i}(N)+\sum_{i=1}^{k+1} B_{i^{'}}(N)
\end{equation}
and hence the critical point can be obtained by demanding that 
$Q_{N}(\sigma)$ is 0.5
as the critical point is reached. B(N) can be written in a column matrix form 
where $B_{i}(N)$ and $B{i'}(N)$ are its (2k+1) components. The set of equations
represented by Eq.(14) can then be written as:
\begin{eqnarray}
B(N)=M.B(N-1)
\end{eqnarray}
M can be identified as a square matrix with $(2k+1)^{2}$ components, all 
independent of N. M consists of the following elements:
$$  \rm first~ row~~   M_{1,i}=0,~~~ M_{1,i^{'}}=p_{1}$$
$$  \rm second~ row~~  M_{1^{'},i}=0,~~~ M_{1^{'},i^{'}}=q_{1}$$
rest of the rows:
$$ M_{j,i}=\delta_{j-1,i} p_{j},~~~   M_{j,i^{'}}=0;$$
                                                                                
$$(i,j > 1)~~~M_{j^{'},i}=\delta_{j-1,i} q_{j},~~~M_{j^{'},i{'}}=0$$
where i=1,...k and $i^{'}=1^{'},2^{'}.......,(k+1)^{'}$.

Eq. (16) can now be written as:
\begin{eqnarray}
B(N)=M^{N}B(1)
\end{eqnarray}
where vector $B(1)$ is given by $B_{i}(1)=\delta_{i,1}p_{1}$, and
$B_{i^{'}}(1)=\delta_{i^{'},1^{'}}q_{1}$

Since the model is based on the $k$-failure concept, the next question that
arises is what an appropriate value for $k$ should be (which fixes the
matrix size). This can be obtained by a simple criterion of convergence:
if results do not vary when matrix size is increased from $k$ to $k+1$, then
that value of $k$ is sufficient for $k$-failure concept. This is equivalent to 
saying that the probability of breaking when the stress on a fiber 
is $\it{(k+1)\sigma}$
is 1 (by the definition of $k$-failure)and hence there is no further change as 
one increases $k$(as the complete failure has already occurred).  

\begin{figure}[h]
\includegraphics[height=2.0in,width=1.5in]{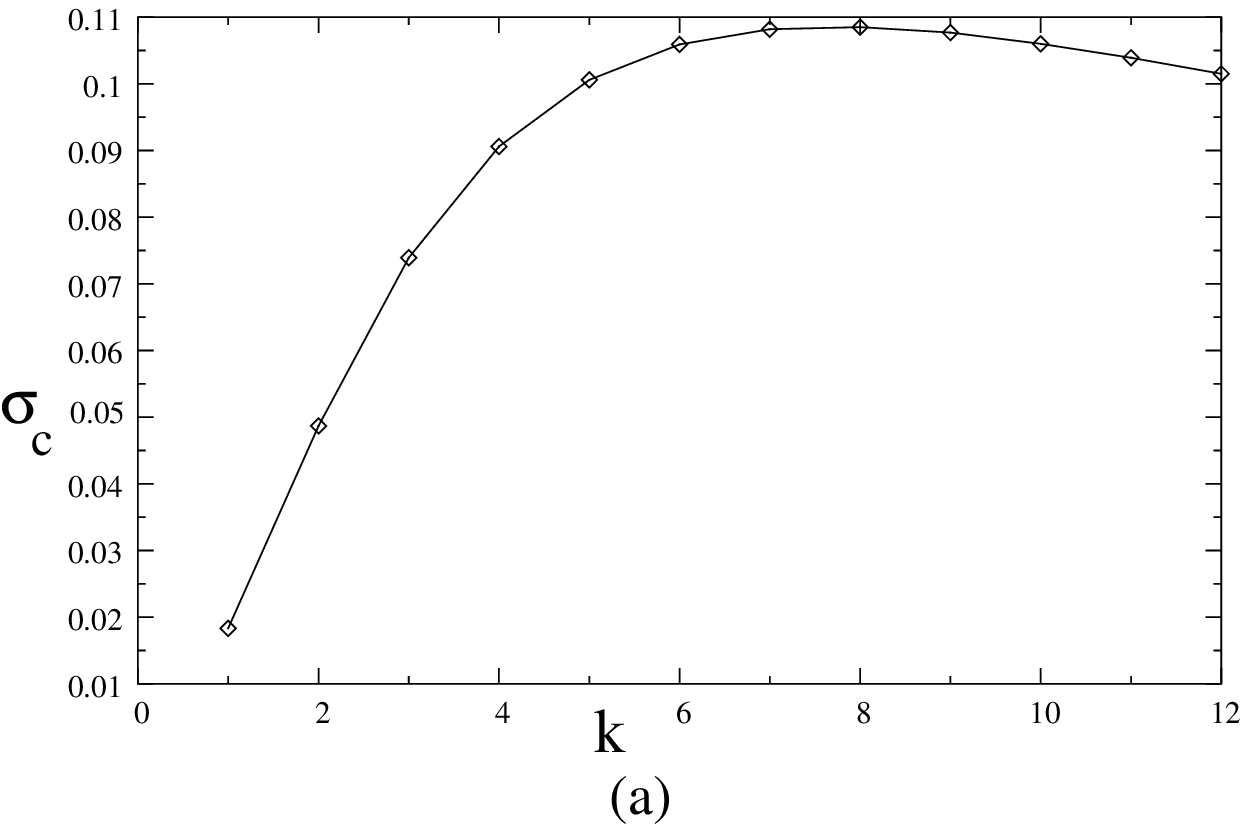}
\includegraphics[height=2.0in,width=1.5in]{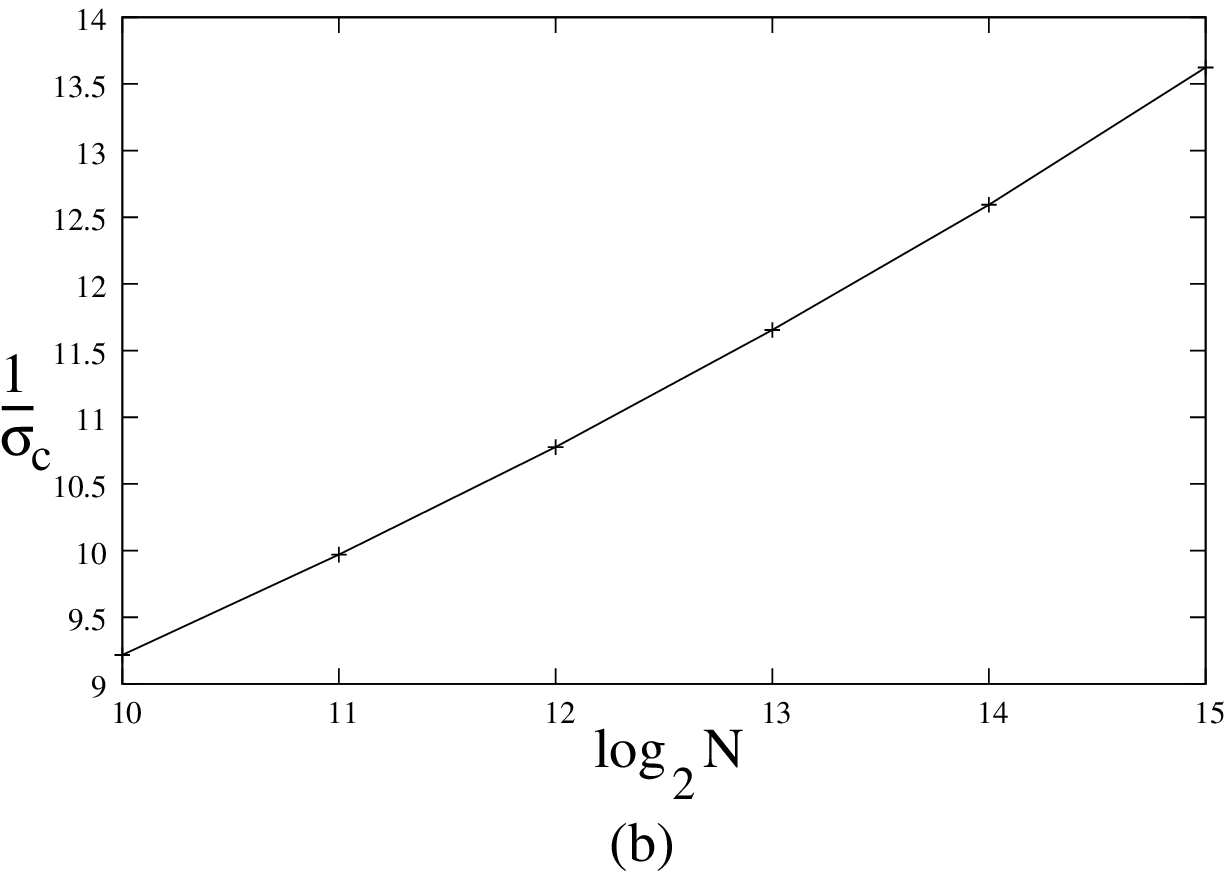}
\caption{The variation of $\sigma_{c}$ with $k$ for UD is shown in Fig~8(a).
~Fig.~ 8(b) shows the variation of $1/\sigma_{c}$ with $\rm log_{2} N$.}
\end{figure}
The above described convergence can be clearly seen if one uses 
WD of threshold strength\cite{Pacheco}.
As mentioned before, the situation is different in the case of UD.
The method used above allows the probability of failure to go beyond 1 if 
UD is used. The  probability that a fiber breaks when the stress is 
$x\sigma$ is $x\sigma$, where $x$ should vary from 1 to $k$. 
It may happen that for some $x$ and $\sigma$, 
the matrix elements become greater than 1 which is clearly unphysical as
each matrix element corresponds to a probability.
It is observed that the value of $\sigma_{c}$ increases with $k$ and after 
attaining an almost constant $\sigma_{c}$, it starts decreasing. This decrease
 actually corresponds to the situation when the matrix elements at some
stage of the calculation is either greater than 1 or less than 0 and hence
 not a saturation but a peak is obtained. Fig.~8(a) shows the variation 
of $\sigma_{c}$ with $k$ where a peak
 is observed. We
have taken $k=8$ for our further calculations. 

In Fig.~8(b), we have shown the variation of 1/$\sigma_{c}$ with 
log$_{2}N$. Clearly it is
a straight line which matches with the previously obtained analytical
and numerical results\cite{Pacheco, Hansen, Smith}. The slope of the line
is 0.81. The critical stress vanishes in the thermodynamic
limit indicating the absence of a critical behaviour in one sided, 
one dimensional RFBM.
\section{Fibers on a random graph}

Construct a random graph having $N$ sites such that each site has exactly 3
neighbours. The algorithm used is as follows: label the sites by integers from 
1 to $N$ where $N$ is even. Connect site $i$ to $i+1$ for all $i$ 
such that site $N$ is 
connected to site 1. We call this connection as direct connection. 
Thus we have a ring of $N$ sites. 
Now connect each site randomly to a unique site other than $i+1$ and $i-1$, 
thus forming $N/2$ pairs of sites. 
This connection is called 
random connection. Each site in such a graph has a coordination number 3. 
In this
construction, all sites are on the same footing (See Fig.~9). The 
tip of the fibers are placed on 
these $N$ sites and are associated with a UD of threshold 
strength. 

\begin{figure}[t]
\includegraphics[height=2.2in,width=2.3in]{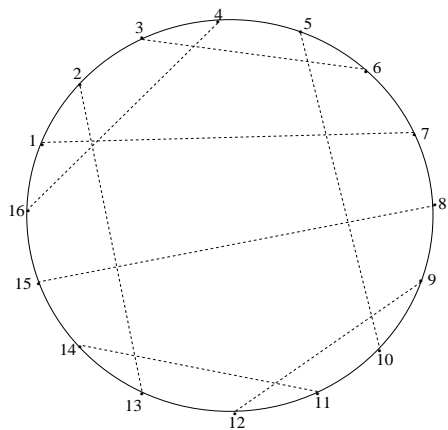}
\caption{An example of a random graph with coordination number 3 and 16 sites}
\end{figure}
\begin{figure}[h]
\includegraphics[height=2.2in,width=2.2in]{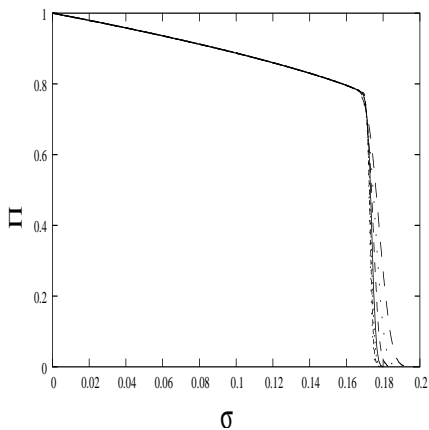}
\caption{The variation of $\pi$ with $\sigma$ for different values of $N$.  
Here, $N=2^{13}, 2^{14}, 2^{15}, 2^{16}, 2^{17},
10^{5},2*10^{5}$. The lower $\sigma_{c}$ corresponds to a higher value of $N$.}
\end{figure}
The system is then exposed to a total load $N\sigma$ 
such that each vertex 
carries a stress $\sigma(i)=\sigma$. If $\sigma_{th}(i)<\sigma(i)$ at
any site $i$, the fiber breaks and its stress is equally distributed to its 
surviving neighbours. The above condition is examined for each site and is 
repeated until there is no more breaking. Increase the total load 
$N\sigma$ to $N(\sigma+d\sigma)$. The new total load is then shared by 
the remaining unbroken fibers and once again the breaking condition is 
checked at each site. To make sure that the total load is conserved, 
if a fiber encounters a situation in which one of its neighbours is broken,
it gives its stress to the next nearest surviving neighbour in that direction.
This rule however is not applicable to the randomly connected fiber $i.e.$,
if the randomly connected neighbour of a broken fiber is already broken, 
the stress then is given only to its directly connected neighbours. Once again,
the load is increased and the process is repeated till complete breakdown of 
the system.
In the numerical simulations, $d\sigma$ is taken to be 0.0001 and the average is taken over 
$10^{4}$ to $10^{5}$ configurations.

Fig.~(10) shows the variation of $\pi$, the fraction of unbroken fibers, with 
stress $\sigma$ for different values of $N$. Clearly, there exist a non-zero 
value of critical stress even for large $N$.

To determine $\sigma_{c}$ precisely, we use the standard finite size 
scaling method \cite{Kim} with the scaling assumption
\begin{equation}
\pi(\sigma,N) = N^{-a}f((\sigma-\sigma_{c})N^{\frac{1}{\nu}})
\end{equation}
where f(x) is the scaling function and the exponent $\nu$ describes the 
divergence of the correlation length $\xi$ near the critical point, i.e., 
$\xi \propto |\sigma-\sigma_{c}|^{-\nu}$. 
If $\pi \propto (\sigma_{c}-\sigma)^{\beta}$, we obtain the relation 
$a=\beta/\nu$. The plots of $\pi N^{a}$with $\sigma$ for different N
crosses through a unique point $\sigma=\sigma_{c}$=0.173 with the exponent 
$a=0.5$ (See Fig.~11). 
\begin{figure}[h]
\includegraphics[height=2.2in,width=2.2in]{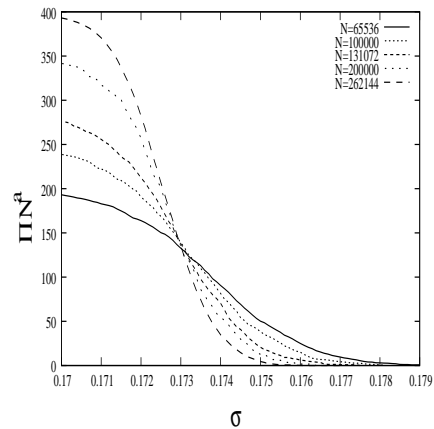}
\caption{Precise determination of $\sigma_{c}$ from the finite size scaling
form (18). The $\sigma_{c}$ at which
the curves intersect is 0.173 }
\end{figure}

With this value of $\sigma_{c}$ and $a$, one can again use the scaling 
relation (18) to determine the  exponent $\nu$ by making the data points to 
collapse to an almost single smooth curve as displayed in Fig.~(12). The 
value of $\nu$ obtained in this manner is equal to 1 which automatically gives the value of the other exponent $\beta$=0.5. This matches with the 
mean field exponents derived analytically\cite{dynamic}. 
A random graph with 
connections as described above is equivalent to a Bethe lattice
with co-ordination number $z$ ($=3$ in present case) for large N \cite{Dhar}.  
It is observed that in this model, 
accurate value of
$\sigma_{c}$(Fig.~11) and data collapse(Fig.~12) obtained by
finite size scaling method is applicable in the limit of large N.
It may be because of the resemblance of the underlying
structure of the system with the Bethe lattice that it
has mean field exponents. This should also be compared to  Ising models
with infinite-range interaction (which corresponds to GLS
 in the present context) 
\cite {Stanley} and with short-range interaction on a Bethe lattice both
showing mean field critical behaviour.

\begin{figure}[h]
\includegraphics[height=2.2in,width=2.2in]{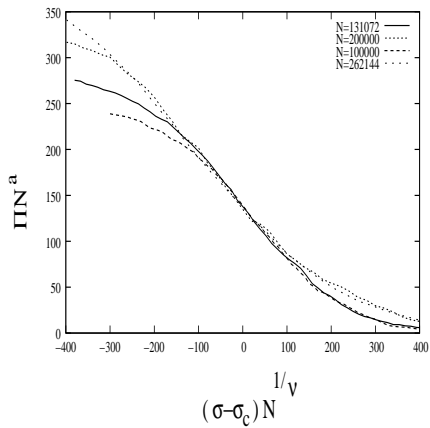}
\caption{Data collapse with $a$=0.5, $\nu=1$}
\end{figure}

\section{Conclusion}
We have introduced a mixed RFBM  with UD in one case and WD in the other
within GLS scheme. The mixed RFBM with UD has a forbidden region 
$\sigma_2 - \sigma_1$ so that the distribution of the threshold strength
is discontinuous though the uniformity is ensured. The width of the forbidden
region tunes the critical stress of the system and thus exhibits a rich
critical behaviour. At the same time, restrictions on the
parameters of the distribution for which the analytical method holds good
are also clearly pointed out.
The burst avalanche distribution
of this type of discontinuous threshold distribution is expected to
be very interesting and that is the subject of our further research. 
To study mixed  RFBM with WD, a quasistatic approach is employed to determine
the critical stress which decreases linearly with the fraction of class A fibers. Exponents obtained in both the above
cases fall under mean field universality class. The most interesting feature
of the mixed models is the tunability of the critical stress with 
the system parameters.
 
The behaviour of critical stress in a one dimensional RFBM with one sided
LLS and strictly UD is also studied using an exact method and the complication
arising due to the uniform nature of distribution is emphasized.
We have also looked at the breakdown phenomena of RFBM on a
random graph with coordination number 3.  The critical stress is obtained 
using the finite size scaling method. The mean field nature of transition
is established in the limit $N \to \infty$.
 
\begin{center}
{\bf ACKNOWLEDGMENTS}
\end{center}
The authors thank P. Bhattacharyya, S. M. Bhattacharjee, A. F. Pacheco and
S. Pradhan for help and suggestions. 
Special thanks to Y. Moreno for helping us 
in writing a code and B. K. Chakrabarti for useful comments.

\end{document}